\documentclass[twocolumn,showpacs,preprintnumbers]{revtex4} 
\usepackage{graphicx}% Include figure files
\usepackage{dcolumn}% Align table columns on decimal point
\usepackage{bm}% bold math
\usepackage{amssymb}
\usepackage{amsmath}
\usepackage{epsfig}    
\usepackage{color}
\def\beq{\begin{equation}}
\def\eeq{\end{equation}}
\def\bea{\begin{array}}
\def\eea{\end{array}}
\def\be{\begin{equation}}
\def\ee{\end{equation}}
\def\ba{\begin{eqnarray}}
\def\ea{\end{eqnarray}}

\def\[{\left[}
\def\]{\right]}
\def\({\left(}
\def\){\right)}

\def\sm0{{\widetilde{m}_0}}

\def\U1em{{U(1)_{\rm em}}}

\def\sq2{\sqrt{2}}

\def\ee{e^+e^-}

\def\End{\end{document}}
\newcommand{\gsim}{\mbox{ \raisebox{-1.0ex}{$\stackrel{\textstyle >}
{\textstyle \sim}$ }}}
\newcommand{\lsim}{\mbox{ \raisebox{-1.0ex}{$\stackrel{\textstyle <}
{\textstyle \sim}$ }}}
\newcommand{\aneq}{\!\!\!&=&\!\!\!} 
  \def\gtsim{\mathrel{\hbox{\raise0.2ex
\hbox{$>$}\kern-0.75em\raise-0.9ex\hbox{$\sim$}}}}
\def\ltsim{\mathrel{\hbox{\raise0.2ex
\hbox{$<$}\kern-0.75em\raise-0.9ex\hbox{$\sim$}}}}
\def\fsl#1{\setbox0=\hbox{$#1$}                 % set a box for #1 
   \dimen0=\wd0                                 % and get its size
   \setbox1=\hbox{/} \dimen1=\wd1               % get size of /
   \ifdim\dimen0>\dimen1                        % #1 is bigger
      \rlap{\hbox to \dimen0{\hfil/\hfil}}      % so center / in box
      #1                                        % and print #1
   \else                                        % / is bigger
      \rlap{\hbox to \dimen1{\hfil$#1$\hfil}}   % so center #1
      /                                         % and print /
   \fi}

\begin{document}                                                              
%\draft
%@twocolumnfalse\endcsname

\title{First-order electroweak phase transition powered by additional F-term
loop effects\\ in an extended supersymmetric Higgs sector}%
\preprint{UT-HET 058, KIAS-P11057, KU-PH-010}
\author{%
{\sc Shinya Kanemura\,$^1$, 
     Eibun Senaha\,$^{2,3}$,
     Tetsuo Shindou\,$^4$}  
     }
\affiliation{
\vspace*{4mm} 
$^1$Department of Physics, University of Toyama, 3190 Gofuku, Toyama 930-8555, Japan\\ 
$^2$ Korea Institute for Advanced Study, School of Physics, 85 Hoegiro, Dongdaemun-gu, 
Seoul 130-722, Korea \\
$^3$ Physics Division, National Center for Theoretical Sciences,
Hsinchu, Taiwan 300 \\
$^4$Division of Liberal Arts, Kogakuin University, 
    1-24-2 Shinjuku, Tokyo 163-8677, Japan
}

\begin{abstract}
 We investigate the one-loop effect of new charged scalar bosons on the
 Higgs potential at finite temperatures in the supersymmetric standard
 model with four Higgs doublet chiral superfields as well as a pair of
 charged singlet chiral superfields.
 In this model, the mass of the lightest Higgs boson $h$ is determined
 only by the D-term in the Higgs potential at the tree-level, 
 while the triple Higgs boson coupling for $hhh$ can receive a significant
 radiative correction due to nondecoupling one-loop contributions
 of the additional charged scalar bosons. 
 We find that the same nondecoupling mechanism can also contribute to
 realize stronger first order electroweak phase transition than that
 in the minimal supersymmetric standard model, which  is definitely 
 required for a successful scenario of electroweak baryogenesis. 
 Therefore, this model can be a new candidate for a model in which  
 the baryon asymmetry of the Universe is explained at the electroweak scale.
 %\hfill~~[\today] }
\end{abstract}

\maketitle

\setcounter{footnote}{0}
\renewcommand{\thefootnote}{\arabic{footnote}}

\section{Introduction}

%[Baryogenesis overview]
It has been confirmed that the numbers of matter and anti-matter in our Universe are asymmetric.
The recent observation suggests that the ratio for baryon-to-photon is
given by  
$n_b/n_\gamma \simeq (5.1-6.5) \times 10^{-10}$
 at the 95 \% CL~\cite{BAU}, where
$n_b$ is the difference in number density between baryons and
anti-baryons and $n_\gamma$ is the number density of photons.  
Understanding the mechanism of generating
the baryon asymmetry of the Universe has
been one of the most important issues
in particle physics and cosmology for a long time.
In order to generate the baryon asymmetry from the baryon symmetric
initial state, the following Sakharov's three conditions have to be
satisfied~\cite{sakharov}:
1) Baryon number nonconservation, 
2) C and CP violation, 
3) Departure from the thermal equilibrium.
%

%[Electroweak baryogenesis]

The electroweak gauge theory can naturally satisfy the above three conditions.
The baryon number nonconservation is realized by the sphaleron process
at high temperature. The C violation can naturally occur in chiral 
gauge theories, and CP violating phases can remain in
the matter sector after rephasing.
Finally, the circumstance of thermal non-equilibrium can
appear if the electroweak phase transition (EWPT) is of strongly first order.
This scenario is often called the electroweak baryogenesis~\cite{B-EW,ewbg}.
The most attractive feature of this scenario would be its testability. 
The scenario is necessarily related to the dynamics of electroweak symmetry
breaking, so that it can be directly tested by measuring the
Higgs boson properties at collider experiments.
%[Higgs search at the LHC]
Currently, Higgs boson searches are underway at the Tevatron and the
LHC. We expect that the physics of electroweak symmetry breaking
will be clarified in near future.
Then the models of electroweak baryogenesis can be experimentally tested.

In the standard model (SM), it is found that the CP violation by the
Cabibbo-Kobayashi-Maskawa matrix is quantitatively
insufficient~\cite{ewbg_sm_cp}. 
In addition, the requirement of sufficiently strong first order EWPT 
predicts the Higgs boson mass to be too small to satisfy 
the constraint from the LEP direct search results~\cite{h-search-LEP}.
Therefore, in order to realize the baryon asymmetry of the Universe
we need to consider an extension of the SM. 

%[Various models for EWBG]
One of the viable models for successful electroweak baryogenesis would be the two Higgs
doublet model (THDM)~\cite{B-EW3}. The extension of the Higgs sector can
introduce additional CP violating phases, and the quantum effect of extra scalar
bosons in the Higgs potential makes it possible to realize 
sufficiently strong first order EWPT without contradicting
the LEP data for the mass of the Higgs boson.
In Ref.~\cite{ewbg-thdm2}, the connection between the first order EWPT
and the triple coupling for the lightest SM-like Higgs boson $h$ (the $hhh$ coupling)
has been clarified. In the model with sufficiently strong first order EWPT,
the $hhh$ coupling constant significantly deviates from the SM
prediction due to the same nondecoupling quantum effects
of additional scalar bosons which make the first order EWPT strong.
Such nondecoupling effects on the $hhh$ coupling constant have been
studied in Ref.~\cite{KOSY}.
The scenario of electroweak baryogenesis by nondecoupling loop effects
of extra bosons has also been applied in a TeV scale model 
where tiny neutrino masses, dark matter and the baryon asymmetry of
the Universe may be simultaneously explained~\cite{Aoki:2008av}.

%[SSUSY]
It would also be attractive to consider the scenario of
electroweak baryogenesis in the model based on supersymmetry (SUSY).
SUSY is a good candidate of new physics, which eliminates the
quadratic divergence in the one-loop calculation of the Higgs boson
mass. The lightest SUSY partner particle in SUSY models with the R
parity can naturally be a candidate for the cold dark matter.    

In the minimal supersymmetric SM (MSSM), there are many studies to realize
the electroweak baryogenesis~\cite{ewbg-mssm,Carena:2008vj,Funakubo:2009eg}. 
Currently, this scenario is highly constrained by the experimental data, especially the LEP Higgs mass bounds, leading to the tension between the lightest Higgs boson mass and the strength of the first order
EWPT. Nevertheless, it is still viable for some specific mass spectrum.
 According to Ref. \cite{Carena:2008vj}, the strong first order EWPT is possible if
$m_h\lsim 127$~GeV and $m_{\tilde{t}_1}\lsim 120$~GeV,
where $h$ is the lightest Higgs boson and $\tilde{t}_1$ is the lightest
stop.  
To satisfy the LEP bound on $m_h$, the soft SUSY breaking mass for the left-handed stop 
should be greater than 6.5~TeV. The most striking feature of this scenario is that
the electroweak vacuum is metastable and the global minimum is a charge-color-breaking vacuum,
where the lifetime of the electroweak
vacuum is found to be longer than the age of the Universe.

The aforementioned tension in the MSSM baryogenesis
can be relaxed by extending the Higgs sector.
One of the possibilities is to add a gauge singlet field into the MSSM.
So far, many studies on the electroweak baryogenesis
have been done in such singlet-extended MSSMs; i.e., 
the Next-to-MSSM~\cite{Funakubo:2005pu}, the nearly MSSM or the minimal
non-MSSM~\cite{EWPT_nMSSM}, the $U(1)'$-extended MSSM~\cite{EWPT_UMSSM},  
the secluded $U(1)'$-extended MSSM~\cite{Chiang:2009fs,Kang:2009rd},
and so on.
In this class of the models, the strong first order EWPT can be induced 
by the trilinear mixing term of the doublet and the singlet fields appearing in the tree-level Higgs potential.
If this is the case, the mass constraints on the lightest Higgs boson
and the light stop would be alleviated significantly.
In the singlet-extended MSSM, however, the vacuum structure is inevitably more complicated
than the MSSM, giving rise to the unrealistic vacua 
in the large portion of the parameter space, especially electroweak baryogenesis-motivated
scenario~\cite{Funakubo:2005pu,Chiang:2009fs}.

%}

In this Letter, we consider how the electroweak phase transition can be
of sufficiently first order due to the nondecoupling effect of additional
scalar bosons in another extended SUSY standard model, where 
a pair of extra  doublet chiral superfields $H_3$ ($Y=-1/2$) and $H_4$ ($Y=+1/2$)
and a pair of charged singlet chiral superfields $\Omega_1$ ($Y=+1$) and
$\Omega_2$ ($Y=-1$) are introduced in addition to the MSSM content~\cite{ksy}.
Since the singlet fields are charged rather than neutral, 
they do not have the vacuum expectation values (VEVs)
as long as the $U(1)_{\rm EM}$ is preserved.
Therefore, at least at the tree level,
there is no vacuum instability caused by the singlet field
as opposed to the neutral singlet-extended MSSM as mentioned above.
On top of this, this model may be motivated by SUSY extensions of
the models with additional charged singlet fields in which
neutrino masses are generated by radiative corrections~\cite{SS-R,Aoki:2008av}.
In the present model, apart from the model in which an additional neutral singlet
chiral superfield is added to the MSSM, there is no tree-level F-term
contribution to the mass of the lightest Higgs boson,
but there can be large one-loop corrections to the triple Higgs boson
coupling due to the additional bosonic loop
contribution~\cite{ksy}.
We note that in the SUSY Higgs sector with four doublet chiral superfields~\cite{aksy}, 
all the quartic interactions in the Higgs potential come from the D-term 
so that there are no large nondecoupling quantum effects on the $hhh$ coupling~\cite{ksy}. 
Similarly to the case of the non-SUSY THDM, these nondecoupling
bosonic loop contributions can also make first order phase transition stronger.
We here show that the EWPT can be of sufficiently
strong first order in this model.

\section{Model}

\begin{table}[t]
\begin{tabular}{|c|cc||ccc|c|}\hline
&Spin $0$&Spin $1/2$&SU(3)&SU(2)&$\mbox{U(1)}_Y$
&$Z_2$\\ \hline
&&&&&&
\\[-3mm]
$Q_i^{}$&$\tilde{q}_L^{}=\begin{pmatrix}\tilde{u}_L^{}\\ \tilde{d}_L^{}\end{pmatrix}$&
$q_L^{}=\begin{pmatrix}u_L^{}\\ d_L^{} \end{pmatrix}$&
$\pmb{3}$&$\pmb{2}$&$+\frac{1}{6}$&$+$\\[3mm]
$U_i^c$&$\tilde{u}_R^*$&$\bar{u}_R^{}$&$\pmb{\bar 3}$&$\pmb{1}$&
$-\frac{2}{3}$&$+$\\[3mm]
$D_i^c$&$\tilde{d}_R^*$&$\bar d_R^{}$&$\pmb{\bar 3}$&$\pmb{1}$&
$+\frac{1}{3}$&$+$\\[3mm]
$L_i^{}$&$\tilde{\ell}_L^{}=\begin{pmatrix}\tilde{\nu}_L^{}\\ \tilde{e}_L^{}\end{pmatrix}$&
$\ell_L^{}=\begin{pmatrix}\nu_L^{}\\ e_L^{} \end{pmatrix}$&
$\pmb{1}$&$\pmb{2}$&$-\frac{1}{2}$&$+$\\[3mm]
$E_i^c$&$\tilde{e}_R^*$&$\bar e_R^{}$&$\pmb{1}$&$\pmb{1}$&
$+1$&$+$\\[1mm] \hline
&&&&&&
\\[-3mm]
$H_1^{}$&$\Phi_1^{}=\begin{pmatrix}\varphi^0_1\\ \varphi^-_1\end{pmatrix}$&
$\tilde{\Phi}_{1L}^{}=\begin{pmatrix}\tilde{\varphi}^0_{1L}\\ \tilde{\varphi}^-_{1L}\end{pmatrix}$
&$\pmb{1}$&$\pmb{2}$&$-\frac{1}{2}$&$+$\\[3mm]
$H_2^{}$&$\Phi_2^{}=\begin{pmatrix}\varphi^+_2\\ \varphi^0_2\end{pmatrix}$&
$\tilde{\Phi}_{2L}^{}=\begin{pmatrix}\tilde{\varphi}^+_{2L}\\ \tilde{\varphi}^0_{2L}\end{pmatrix}$
&$\pmb{1}$&$\pmb{2}$&$+\frac{1}{2}$&$+$\\[3mm]\hline
$H_3^{}$&$\Phi_3^{}=\begin{pmatrix}\varphi^0_3\\ \varphi^-_3\end{pmatrix}$&
$\tilde{\Phi}_{3L}^{}=\begin{pmatrix}\tilde{\varphi}^0_{3L}\\ \tilde{\varphi}^-_{3L}\end{pmatrix}$
&$\pmb{1}$&$\pmb{2}$&$-\frac{1}{2}$&$-$\\[3mm]
$H_4^{}$&$\Phi_4^{}=\begin{pmatrix}\varphi^+_4\\ \varphi^0_4\end{pmatrix}$&
$\tilde{\Phi}_{4L}^{}=\begin{pmatrix}\tilde{\varphi}^+_{4L}\\ \tilde{\varphi}^0_{4L}\end{pmatrix}$
&$\pmb{1}$&$\pmb{2}$&$+\frac{1}{2}$&$-$\\[3mm]
$\Omega_1^{}$&$\omega_1^+$&
$\overline{\tilde{\omega}}_1^{-}$
&$\pmb{1}$&$\pmb{1}$&$+1$&$-$\\[3mm]
$\Omega_2^{}$&$\omega_2^{-}$&
$\tilde{\omega}_{2}^{-}$
&$\pmb{1}$&$\pmb{1}$&$-1$&$-$\\[3mm]
\hline
\end{tabular}
 \caption{Property of chiral superfields (and their component fields)
 under the symmetries of the model.}
\end{table}

We consider the model with the chiral superfields shown in Table~1~\cite{ksy}.
The symmetries of the model are standard gauge symmetries.
In addition, we impose a discrete $Z_2$ symmetry for simplicity. 
Although the $Z_2$ symmetry is not essential for our discussion,
the symmetry works for avoiding the flavor changing neutral
current at the tree level~\cite{fcnc,barger,typeX,aksy}. 
Furthermore, we assume that there is the R parity in our model.

%[Lagrangian]

The superpotential is given by 
\begin{align}
 W &=(y_u)^{ij} U_i^c H_2\cdot Q_j+(y_d)^{ij} D_i^c H_1\cdot Q_j
 +(y_e)^{ij}E^c_iH_1\cdot L_j \nonumber \\
 &+ \lambda_1\Omega_1 H_1\cdot H_3 
 + \lambda_2\Omega_2 H_2\cdot H_4 \nonumber\\
 &-\mu H_1\cdot H_2 -\mu^\prime H_3\cdot H_4 - \mu_\Omega \Omega_1 \Omega_2.
 \end{align}
The soft-SUSY-breaking terms are given by 
\begin{align}
 {\mathcal L}_{\rm soft} &=
 -\frac{1}{2}(M_1 \tilde{B}\tilde{B}
             +M_2 \tilde{W}\tilde{W}
             +M_3 \tilde{G}\tilde{G})\nonumber\\
 &-\left\{
 (\tilde{M}_{\tilde{q}}^2)_{ij} \tilde{q}_{Li}^\dagger \tilde{q}_{Lj}  +
 (\tilde{M}_{\tilde{u}}^2)_{ij} \tilde{u}_{Ri}^\ast \tilde{u}_{Rj}
  +\right. \nonumber\\
 &\left.
 (\tilde{M}_{\tilde{d}}^2)_{ij} \tilde{d}_{Ri}^\ast \tilde{d}_{Rj}+
 (\tilde{M}_{\tilde{\ell}}^2)_{ij} \tilde{\ell}_{Li}^\dagger \tilde{\ell}_{Lj}  +
 (\tilde{M}_{\tilde{e}}^2)_{ij} \tilde{e}_{Ri}^\ast \tilde{e}_{Rj} 
 \right\} \nonumber\\
 &-\left\{
   \tilde{M}_{H_1}^2\Phi_1^\dagger\Phi_1
+  \tilde{M}_{H_2}^2\Phi_2^\dagger\Phi_2
+  \tilde{M}_{H_3}^2\Phi_3^\dagger\Phi_3
  +\right. \nonumber\\
 &\left.
+  \tilde{M}_{H_4}^2\Phi_4^\dagger\Phi_4
+  \tilde{M}_{+}^2 \omega_1^+\omega_1^-
+  \tilde{M}_{-}^2 \omega_2^+\omega_2^-
 \right\} \nonumber\\
&- \left\{
 (A_u)^{ij} \tilde{u}_{Ri}^\ast \Phi_2 \cdot \tilde{q}_{Lj}
+(A_d)^{ij} \tilde{d}_{Ri}^\ast \Phi_1 \cdot \tilde{q}_{Lj}
  +\right. \nonumber\\
 &\left.
+(A_e)^{ij} \tilde{e}_{Ri}^\ast \Phi_1 \cdot \tilde{\ell}_{Lj}
 \right.\nonumber\\
&
\left.
+(A_1) \omega_1^+ \Phi_1 \cdot \Phi_3 
+(A_2) \omega_2^- \Phi_2 \cdot \Phi_4 + {\rm h.c.} 
 \right\}. 
\end{align}
From $W$ and $\mathcal{L}_{\text{soft}}^{}$, the Lagrangian is constructed as 
\begin{align}
\mathcal{L}=&
\mathcal{L}_{\text{kinetic}}
+\mathcal{L}_{\text{gauge--matter}}\nonumber\\
&-\left(\frac{1}{2}\frac{\partial^2 W}{\partial \varphi_i\partial\varphi_j}
\psi_{Li}\cdot\psi_{Lj}+h.c.\right)\nonumber\\
& -\left|\frac{\partial W}{\partial \varphi_i}\right|^2-\frac{1}{2}(g_a)^2
(\varphi^*_{\alpha}T^a_{\alpha\beta}\varphi_{\beta})^2+\mathcal{L}_{\rm soft}
\;,
\end{align}
where $\varphi_i$ and $\psi_{L i}$ are respectively
scalar and fermion components of chiral superfields, and    
$T_{\alpha\beta}^a$ and $g_a$ represent generator matrices for
the gauge symmetries and corresponding gauge coupling constants.

The scalar component doublet fields $\Phi_i$ are parameterized as 
\begin{align}
 \Phi_{1,3} =& \left[\begin{array}{c}
      \frac{1}{\sqrt{2}}(\varphi_{1,3} + h_{1,3}+ i a_{1,3}) \\
      \phi_{1,3}^-
                    \end{array} \right], \nonumber \\ 
 \Phi_{2,4} =& \left[\begin{array}{c}
      \phi_{2,4}^+ \\
      \frac{1}{\sqrt{2}}(\varphi_{2,4} + h_{2,4}+ i a_{2,4}) 
                    \end{array} \right], 
 \end{align}
where $\varphi_{i}$ are classical expectation values,
$h_i$ are CP-even, $a_i$ are CP-odd and $\phi_i^\pm$ are charged
scalar states.
where $\varphi_{i}$ are classical expectation values,
$h_i$ are CP-even, $a_i$ are CP-odd and $\phi_i^\pm$ are charged
scalar states.
We use the effective potential method to explore the Higgs sector.
At the tree level, the effective potential for the Higgs fields is
given by 
\begin{eqnarray}
\lefteqn{V_0(\varphi_1,\varphi_2, \varphi_3, \varphi_4)}\nonumber \\
\aneq \sum_{a=1}^4\frac{1}{2}\bar{m}_a^2\varphi_a^2
	+\frac{1}{2}(B\mu \varphi_1\varphi_2+B'\mu' \varphi_3\varphi_4+{\rm h.c.}) \nonumber\\
&&	+\frac{g^2+g'^2}{32}(\varphi_1^2-\varphi_2^2+\varphi_3^2-\varphi_4^2)^2.
\end{eqnarray}
Using the effective potential, the vacuum is determined 
by the stationary condition as
\begin{align}
\left. \frac{\partial V_{\rm eff}}{\partial \varphi_i}
 \right|_{\langle\varphi_i\rangle=v_i} =0.
 \label{VeffVac}
\end{align}
We assume that the $Z_2$ odd Higgs bosons do not have the VEVs ($v_3 = v_4 = 0$) for simplicity, 
and we set
$\sqrt{v_1^2+v_2^2} \equiv v$ ($\simeq 246$~GeV) and introduce $\tan\beta=v_2/v_1$. 
At the tree level, $v_3 = v_4 = 0$ is guaranteed by requiring the nonnegative eigenvalues of 
$(\partial^2 V_0/\partial \varphi_i\partial \varphi_j)_{\varphi_{i,j}=0}~(i,j=3,4)$, i.e., 
\begin{eqnarray}
\bar{m}_3^2\bar{m}_4^2-B'^2\mu'^2\geq 0,\quad \bar{m}_3^2+\bar{m}_4^2\geq0.
\label{Z2sym}
\end{eqnarray}
In the following, we exclusively focus on the $(\varphi_1, \varphi_2)$ space.

For the $Z_2$ even scalar states, after the symmetry breaking
we have five physical states as in the MSSM; i.e.,
two CP-even $h$ and $H$, a CP-odd $A$ and a pair of charged $H^\pm$
scalar bosons.
The tree level mass formulae for these scalar states coincide with those
in the MSSM. 

The mass eigenstates for the $Z_2$ odd charged scalar states
$\phi^{\prime\pm}_{1}$ and $\phi^{\prime\pm}_2$
are obtained by diagonalizing the component fields of
doublet scalar fields $\Phi_3$ and $\Phi_4$, and 
$\Omega^{\pm}_{1}$ and $\Omega^{\pm}_{2}$ from
the charged singlet scalar fields $\omega_1$ and $\omega_2$.
Their field dependent masses are given by 
\begin{align}
\overline{m}_{\phi_{1,2}^{\prime\pm}}^{2} &= \frac{1}{2}
  \left[ \overline{m}_3^2 + \overline{m}_4^2
 + \frac{1}{2} \left(|\lambda_1|^2\varphi_1^2+|\lambda_2|^2\varphi_2^2
 \right)
% \right.\nonumber\\
%& \left.
 \mp
 \sqrt{D_{\phi^{\prime\pm}}}\right], \nonumber \\
 \overline{m}_{\Omega_{1,2}^{\pm}}^{2} &= \frac{1}{2}
  \left[ \overline{m}_+^2 + \overline{m}_-^2
 + \frac{1}{2} \left(|\lambda_1|^2\varphi_1^2+|\lambda_2|^2\varphi_2^2
 \right)
% \right.\nonumber\\
%& \left.
 \mp
 \sqrt{D_{\Omega^{\pm}}}\right], 
\end{align} 
where
\begin{align}
 \overline{D}_{\phi^{\prime\pm}} =&
\left(\overline{m}_3^2 - \overline{m}_4^2
 + \frac{1}{2} \left(|\lambda_1|^2\varphi_1^2-|\lambda_2|^2\varphi_2^2 \right)\right.\nonumber\\
& \left. -\frac{g^2-g^{\prime 2}}{4}(\varphi_1^2-\varphi_2^2)\right)^2 + 4 |B^\prime|^2
 |\mu^\prime|^2, \\
  \overline{D}_{\Omega^{\pm}} =&
\left(\overline{m}_+^2 - \overline{m}_-^2
 + \frac{1}{2} \left(|\lambda_1|^2\varphi_1^2-|\lambda_2|^2\varphi_2^2 \right)\right.\nonumber\\
& \left. -\frac{g^{\prime 2}}{2}(\varphi_1^2-\varphi_2^2)\right)^2 + 4 |B_\Omega|^2 |\mu_\Omega|^2.
\end{align}
Mass parameters in the above formulas are $\overline{m}_{3,4}^2=\tilde{M}_{H_{3,4}}^2+|\mu^{\prime}|^2$ 
and $\overline{m}_{+,-}^2=\tilde{M}_{+,-}^2+|\mu_{\Omega}|^2$.
The physical masses can be obtained by replacing $\varphi_i$ by
$v_i$, the values at the vacuum. 
On the other hand, masses of the additional $Z_2$ odd neutral scalar
bosons do not receive the contributions from the $|\lambda_i|^2$-terms,
and only receive those from the D-term and the soft-SUSY-breaking
terms. Their loop effect on the effective potential are small, so that
  we neglect them in our later calculations.

The field dependent masses of the $Z_2$ odd charginos are given by 
\begin{align}
\bar{m}_{\tilde{\chi}_{1,2}^{\prime \pm}}^2=\frac{1}{2}&\left[|\mu^{\prime}|^2+|\mu_{\Omega}|^2\phantom{\frac{1}{2}}\right.\nonumber\\
&\left.+\frac{1}{2}\left(|\lambda_1|^2|\varphi_1|^2+|\lambda_2|^2|\varphi_2|^2\right)
\mp\sqrt{\bar{D}_{\tilde{\chi}}}\right]\;,
\end{align}
where 
\begin{align}
\bar{D}_{\tilde{\chi}}=&\left(|\mu^{\prime}|^2-|\mu_{\Omega}|^2+
\frac{1}{2}\left(|\lambda_1|^2|\varphi_1|^2-|\lambda_2|^2|\varphi_2|^2\right)\right)^2 \nonumber\\
&+2|\lambda_1^*\mu_{\Omega}\varphi_1+\lambda_2\mu^{\prime *}\varphi_2|^2\;.
\end{align}
The physical masses can be obtained by replacing $\varphi_i$ by $v_i$.
 Notice that the masses do not vanish even when
the invariant mass parameters  $\mu^\prime$ and $\mu_\Omega$ are taken
to be zero.

Since it is known that radiative corrections on the Higgs sector
are very important to study the EWPT,
we here focus on the one-loop contribution.
The vacuum at the one-loop level is also determined from Eq.~(\ref{VeffVac}) with
the one-loop corrected effective potential.
The one-loop correction to the effective potential at zero temperature is 
given by
\begin{eqnarray}
V_1(\varphi_1,\varphi_2)
\aneq \sum_ic_i\frac{\bar{m}_i^2}{64\pi^2}
 \left(
 	\ln\frac{\bar{m}_i^2}{M^2}-\frac{3}{2}
 \right),
\end{eqnarray}
where $V_1$ is regularized in the $\overline{\rm DR}$-scheme,
$c_i$ is the degrees of freedom of the species $i$, $M$ is a renormalization scale
which will be set on $m^{\rm pole}_t$.

For the zero temperature $T=0$, the one-loop corrected
mass matrix for the CP even neutral bosons
can be calculated from the effective potential.
We here consider the simple case such that $B^{\prime}=B_{\Omega}=\mu^{\prime}=\mu_{\Omega}=0$
in order to switch off the mixing effects.
By using the effective potential method, the MSSM-like CP even Higgs boson mass matrix 
$(M_h^2)_{ij}$ with the leading $\lambda_{1,2}^4$ contributions is given as
\begin{align}
 (M_h^2)_{11} =&
 m_Z^2c_{\beta}^2-B\mu\tan\beta+\frac{\lambda_1^4 v^2c_{\beta}^2}{16\pi^2}
\ln
 \frac{m_{\Omega_2^{\pm}}^2m_{\Phi_2^{\prime\pm}}^2}{m_{\tilde{\chi}'^{\pm}_1}^4},
 \nonumber \\
 (M_h^2)_{22} =& m_Z^2s_{\beta}^2-B\mu\cot\beta+\frac{\lambda_2^4 v^22_{\beta}^2}{16\pi^2}
\ln
 \frac{m_{\Omega_1^{\pm}}^2m_{\Phi_1^{\prime\pm}}^2}{m_{\tilde{\chi}'^{\pm}_2}^4},
 \nonumber\\
 (M_h^2)_{12} =& (M_h^2)_{21} = B\mu - m_Z^2c_{\beta}s_{\beta},    
 \end{align}
where $1,2$ are labeled as 
\begin{align}
m_{\Phi_1^{\prime \pm}}^2=&
\bar{m}_4^2-\frac{m_Z^2-2m_W^2}{2}c_{2\beta}+\frac{\lambda_2^2s_{\beta}^2}{2}v^2\;,\nonumber\\
m_{\Phi_2^{\prime \pm}}^2=&
\bar{m}_3^2+\frac{m_Z^2-2m_W^2}{2}c_{2\beta}+\frac{\lambda_1^2c_{\beta}^2}{2}v^2\;,\nonumber\\
m_{\Omega_1^{\pm}}^2=&
\bar{m}_-^2+(m_Z^2-m_W^2)c_{2\beta}+\frac{\lambda_2^2s_{\beta}^2}{2}v^2\;,\nonumber\\
m_{\Omega_2^{\pm}}^2=&
\bar{m}_+^2-(m_Z^2-m_W^2)c_{2\beta}+\frac{\lambda_1^2c_{\beta}^2}{2}v^2\;,\nonumber\\
m_{\tilde{\chi}'^{\pm}_1}^2=&\frac{\lambda_1^2v^2c_{\beta}^2}{2}\;,\nonumber\\
m_{\tilde{\chi}'^{\pm}_2}^2=&\frac{\lambda_2^2v^2s_{\beta}^2}{2}\;.
\end{align}
The renormalized mass of the lightest Higgs boson $h$ is then calculated for $m_A \gg m_Z$ as 
\begin{align}
  m_h^2 &\simeq m_Z^2 \cos^22\beta + \mbox{(MSSM-loop)} \nonumber\\
        &
 + \frac{\lambda_1^4v^2 c_\beta^4}{16\pi^2}
    \ln \frac{m_{\Omega_2^\pm}^2 m_{\Phi_2^{\prime\pm}}^2}{m_{\bar{\chi}'^\pm_1}^4}  
 + \frac{\lambda_2^4v^2 s_\beta^4}{16\pi^2}
    \ln \frac{m_{\Omega_1^\pm}^2 m_{\Phi_1^{\prime\pm}}^2}{m_{\bar{\chi}'^\pm_2}^4},
 \end{align}
at the leading $\lambda_{1,2}^4$ contributions,
where the one-loop contribution in the MSSM is mainly from the top and stop loop diagram\cite{mh-MSSM}.

\begin{figure}[t]
\begin{center}
   \epsfig{file=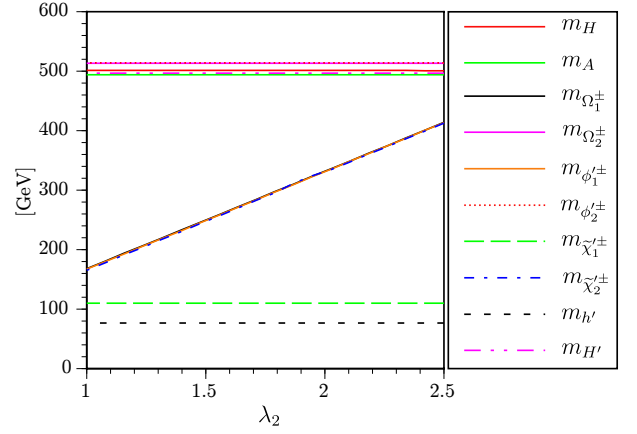,width=8.cm}
\caption{
	The masses of the $Z_2$-odd scalars and 
	$Z_2$-odd charginos as a function of $\lambda_2$. 
	The masses of heavy $Z_2$-even neutral scalars $H$ and $A$
	are also plotted.
	Model parameters are taken as them shown in Eq.~(\ref{Param}).
	The $X_t$ dependence of these mass spectrum is negligible.}
	\label{fig:Z2oddmass}
	\end{center}
\end{figure}
%--------------------------------------  
%--------------------------------------  
\begin{figure}[t]
\center.
\includegraphics[width=6cm]{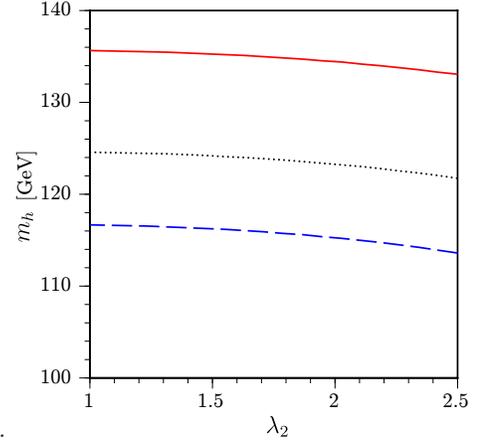}
\caption{The $Z_2$-even lightest Higgs boson mass as a function of $\lambda_2$.
From the top to the bottom, $X_t/\tilde{M}_{\tilde{q}}=2.0, 1.2$ and 0.6.}
\label{fig:mh}
\end{figure}
%-------------------------------------- 
Now we quantify the magnitude of the radiative corrections of the $Z_2$-odd particles on $m_h$.
The input parameters are fixed as follows.
\begin{eqnarray}
{\rm Tree}:&&\!\!\! \tan\beta=3,~m_{H^\pm}=500~{\rm GeV}; \nonumber\\
\mbox{1-loop (MSSM)}:
&&\!\!\! \tilde{M}_{\tilde{q}}=\tilde{M}_{\tilde{b}}=\tilde{M}_{\tilde{t}}=1000~{\rm GeV},\nonumber\\
&&\!\!\! \mu=M_2=2M_1=200~{\rm GeV},\nonumber\\
&&\!\!\! A_t=A_b=X_t+\mu/\tan\beta; \nonumber\\
\mbox{1-loop ($\Phi_{1,2}^{\prime\pm}$, $\Omega$)}:
&&\!\!\! \lambda_1 = 2, \mu'=\mu_\Omega=B_\Omega=B'=0,\nonumber\\
&&\!\!\! \overline{m}_+^2=\overline{m}_3^2=(500\hspace{2mm} {\rm GeV})^2,\nonumber\\   
&&\!\!\! \overline{m}_-^2=\overline{m}_4^2=(50\hspace{2mm} {\rm GeV})^2.
\label{Param}
\end{eqnarray}
We note that $m_{\Phi_1^{\prime\pm}} < m_{\Phi_2^{\prime\pm}}$ and
$m_{\Omega_1^{\pm}} < m_{\Omega_2^{\pm}}$ in this case.  
The mass spectrum of extra $Z_2$ odd charged scalars and charginos
is displayed in Fig.~\ref{fig:Z2oddmass}.
On this parameter set, $m_{\Phi_1^{\prime \pm}}^2$, $m_{\Omega_1^{\pm}}^2$ and 
$m_{\tilde{\chi}'^{\pm}_2}^2$ get a significant contribution from $\lambda_2$.
Then their masses become larger for the greater value of $\lambda_2$.
Since the mass parameters $\bar{m}_4^2$ and $\bar{m}_-^2$ are taken to
be small, large 
mass values of $m_{\Phi_1^{\prime \pm}}$ and $m_{\Omega_1^{\pm}}$
 yield the large nondecoupling effects which can make the EWPT strongly first order.

Fig.~\ref{fig:mh} shows the predicted value of
$m_h$ as a function of $\lambda_2$
varying $X_t/\tilde{M}_{\tilde{q}}=2.0, 1.2$ and $0.6$ from the top to the bottom.
We can see that $m_h$ monotonically decreases as $\lambda_2$ increases, which is 
in contrast with the top/stop loop effects.

The coupling constants $\lambda_1$ and $\lambda_2$ are
free parameters of the model. Its magnitude, however,
is bounded from above by the condition that there is no
Landau pole below the given cutoff scale $\Lambda$.
As we are interested in the model where the first order
EWPT is sufficiently strong, we allow rather
larger values for these coupling constants, and do not
require that the model holds until the grand unification scale. 
A simple renormalization group equation analysis tells us that
for assuming $\Lambda = 2$~TeV, $10$~TeV or $10^{2}$~TeV,
the coupling constant can be taken to be at most $\lambda_2 \sim 2.5$, $2.0$
or $1.5$, respectively. Above the cutoff scale $\Lambda$,
the model may be replaced by a strongly coupled supersymmetric theory
with UV completion
as described by the scenario such as in the fat Higgs model~\cite{fat-higgs}.

\section{Electroweak Phase Transition}
 
The nonzero temperature effective potential is 
\begin{align}
V_1(\varphi_1,\varphi_2;T) = \sum_ic_i\frac{T^4}{2\pi^2}I_{B,F}\left(\frac{\bar{m}_i^2}{T^2}\right),
 \end{align}
where $B(F)$ refer to boson (fermion) and $I_{B,F}$ take the form
\begin{eqnarray}
I_{B,F}(a^2) = \int_0^\infty dx~x^2\ln\Big(1\mp e^{-\sqrt{x^2+a^2}}\Big).\label{IBF}
\end{eqnarray}
Since the minimum search using $I_{B,F}$ is rather time-consuming, 
we will alternatively use the fitting functions of them that are employed in Ref.~\cite{Funakubo:2009eg}. 
More explicitly,
\begin{eqnarray}
\tilde{I}_{B,F}(a^2)=e^{-a}\sum^N_{n=0}c^{b,f}_na^n,
\label{V1_fit}
\end{eqnarray}
are used, where $c^{b,f}_n$ are determined by the least square method.  For $N=40$, $|I_{B,F}(a^2)-\tilde{I}_{B,F}(a^2)|<10^{-6}$ for any $a$, which is sufficient in our investigation. 
Since the nonzero modes of the thermal effective potential give $T^2$ corrections to the 2-point self
energy at high temperatures, we will resum them to make the perturbative analysis more reliable~\cite{Parwani:1991gq}.
 
For an electroweak baryogenesis scenario
to be successful, the sphaleron rate in the broken phase 
should be smaller than the Hubble constant. Conventionally, this condition is translated into 
\begin{eqnarray} 
\frac{v_C}{T_C}=\frac{\sqrt{v_1^2(T_C)+v_2^2(T_C)}}{T_C}\gsim\zeta,
\label{sph_dec}
\end{eqnarray}
where $T_C$ is the critical temperature, $v_C$ is the Higgs VEV at $T_C$, and
$\zeta$ is a $\mathcal{O}(1)$ parameter.
To obtain $\zeta$ within a better accuracy, 
the sphaleron energy and zero-mode factors of the fluctuations around the sphaleron 
must be evaluated. 
In the SM, the sphaleron energy is simply a function of the Higgs boson mass.
As the Higgs boson becomes heavier, the sphaleron energy gets larger as well~\cite{Klinkhamer:1984di},
leading to the smaller $\zeta$~\cite{Arnold:1987mh}. 
In this model, on the other hand, $\zeta$ depends on more parameters.
For simplicity, we here take $\zeta=1$, which is often adopted as a rough criterion in the literature.

In our analysis, $T_C$ is defined as the temperature at which the effective potential has the
two degenerate minima. We search for $T_C$ by minimizing 
\begin{eqnarray}
V_{\rm eff}(\varphi_1, \varphi_2; T) = V_0(\varphi_1, \varphi_2)
+V_1(\varphi_1, \varphi_2)+V_1(\varphi_1, \varphi_2; T), \nonumber \\
\end{eqnarray}
where the field-dependent masses are modified by adding thermal
corrections, 
and Eq.~(\ref{V1_fit}) is used in $V_1(\varphi_1, \varphi_2; T)$.

%--------------------------------------  
\begin{figure}[t] 
\begin{center}
   \epsfig{file=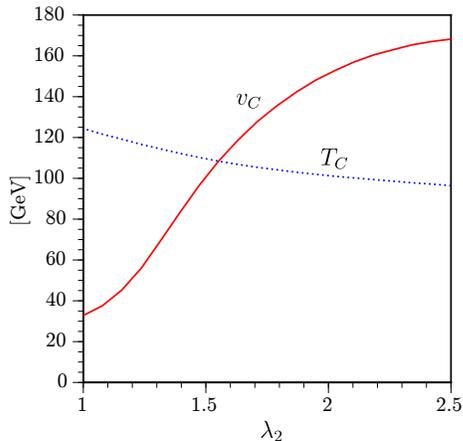,width=6.cm}
\caption{$v_C$ and $T_C$ vs. $\lambda_2$ with $X_t/\tilde{M}_{\tilde{q}} = 0.6$. 
The other input parameters are the same as in the 
Fig.~\ref{fig:mh}. The sphaleron decoupling condition (\ref{sph_dec}) can be satisfied for $\lambda_2\gtsim 1.6$.}
 \label{fig:PT_light}
\end{center}
\end{figure}
%--------------------------------------  
%--------------------------------------  
\begin{figure}[t]
\begin{center}
   \epsfig{file=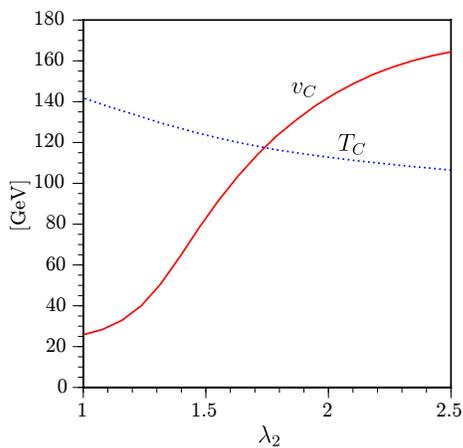,width=6.cm}
\caption{$v_C$ and $T_C$ vs. $\lambda_2$ with $X_t/\tilde{M}_{\tilde{q}} = 2.0$. 
The other input parameters are the same as in the 
Fig.~\ref{fig:mh}. The sphaleron decoupling condition (\ref{sph_dec}) can be satisfied for $\lambda_2\gtsim 1.8$.}
\label{fig:PT_heavy}  
\end{center}
\end{figure}
%--------------------------------------  

In Fig.~\ref{fig:PT_light} $v_C$ and $T_C$ are plotted as a function of $\lambda_2$
in the light $h$ scenario $(X_t/\tilde{M}_{\tilde{q}}=0.6)$: see Fig.~\ref{fig:mh}.  
The sphaleron decoupling condition (\ref{sph_dec}) can be fulfilled for $\lambda_2\gtsim 1.6$ 
due to the nondecoupling effects coming from $\phi'^\pm_1$ and $\Omega_1^\pm$.

We also evaluate $v_C$ and $T_C$ in the heavy $h$ scenario $(X_t/\tilde{M}_{\tilde{q}} = 2.0)$ as shown in Fig.~\ref{fig:PT_heavy}.
The sphaleron decoupling condition can be satisfied for $\lambda_2\gtsim1.8$.
Though the parameter region is a bit narrower than the light Higgs scenario, 
the lightest Higgs boson mass as large as 134~GeV is still consistent with 
the decoupling condition.

 \section{Discussions}

 The several comments on the current analysis are in order.
\begin{itemize}
\item
According to a study of the sphaleron decoupling condition in the MSSM, it is found that
$\zeta\simeq 1.4$~\cite{Funakubo:2009eg} which is 40\% stronger than one we impose in
our analysis. The similar value may be obtained in this model as well.
It should be emphasized, however,  that even if we take $\zeta=1.4$ for the sphaleron decoupling condition, a feasible region exists for the relatively large $\lambda_2$, for example, 
$\lambda_2\gtsim2.2$ even in the heavy $h$ case.
The cutoff scale $\Lambda$ is rather low in this case but still it is
     around the multi-TeV scale. 
     
\item 
In this model, the light stop scenario is one of the options for the
      successful electroweak baryogenesis.
Same as the scenario in the MSSM, the strength of the first order EWPT can get enhanced
if the (almost) right-handed stop is lighter than the top quark, enlarging the possible region.
\item
Similar to the usual MSSM baryogenesis scenario, the charginos and/or the neutralinos
can play an essential role in generating the CP violating sources as needed for the bias of the
chiral charge densities around the Higgs bubble walls. 
In addition to this, the $Z_2$ odd charginos $\widetilde{\chi}'^\pm_{1,2}$ may also do the job
for the successful baryogenesis.

\item
Since $A_{1,2}$ and $\lambda_{1,2}\mu_\Omega$ are small in our parameter choices, 
the charge breaking does not occur at the tree level.
In addition, the $Z_2$ symmetry is not broken spontaneously 
at the tree level, because $\bar{m}_3^2$, $\bar{m}_4^2$ and $B'\mu'$ we take here
satisfy Eq.~(\ref{Z2sym}).
The potential analysis beyond the tree level is out of scope in this Letter.
It will be our future task.

\item
 If the $Z_2$ symmetry is exact and unbroken
 after the electroweak symmetry breaking, the lightest $Z_2$ odd particle
 in our model can be a candidate of cold dark matter if it is electrically
 neutral, in addition to the lightest supersymmetric particle.  
 If one of the extra neutral scalar bosons is the lightest $Z_2$ field,
 its phenomenological property and experimental constraints would be
 similar to those for the supersymmetric extension of the inert doublet
     model~\cite{Barbieri:2006dq}.
 A neutralino from the extra doublets may also be a candidate
 for dark matter.  

\item
Finally, we comment on the phenomenological predictions of this model.
First of all, the nondecoupling effect of the extra $Z_2$ odd charged scalar bosons
on the finite temperature effective potential is an essentially important
feature of our scenario in order to realize strong first order phase transition.
The same physics affects the triple Higgs boson coupling with
a large deviation from the SM (MSSM) prediction 
as discussed in Ref.~\cite{ewbg-thdm2},  Such deviation in the
triple Higgs boson coupling can be      15-70 \% \cite{KOSY,ksy}, and we expect
that they can be measured at the future linear collider such as
the ILC or the CLIC.
Second, in our model, in order to realize the nondecoupling
effect large, the invariant parameters $\mu^\prime$ and $\mu_{\Omega}$
are taken to be small. Consequently, the masses of extra charginos
are relatively as light as 100-300~GeV. 
\end{itemize}

\section{Conclusions}

 We have discussed the one-loop effect of new charged scalar bosons on the
 Higgs potential at finite temperatures in the supersymmetric standard
 model with four Higgs doublet chiral superfields as well as a pair of
 charged singlet chiral superfields.
 We have found that the nondecoupling loop effects of additional charged
 scalar bosons can make first order EWPT strong enough
 to realize successful electroweak baryogenesis. 
 We, therefore, conclude that this model can be a new good candidate
 for a successful model where the baryon asymmetry of the Universe is
 explained at the electroweak scale.
 The detailed analysis for the collider phenomenology will be shown elsewhere.
\\

 \noindent
{\it Acknowledgments}

The authors would like to thank Mayumi Aoki and Kei Yagyu for useful discussions.
This work was supported in part by
Grant-in-Aid for Scientific Research (A) No.~22244031~[S.K.],
Grant-in-Aid for Scientific Research in Priority
Areas No.~22011007~[T.S.], and 
Grant-in-Aid for Scientific Research for Innovative Areas
Nos.~23104006~[S.K.] and 23104011~[T.S.].

%\vspace*{-4mm}

\end{document}